\documentstyle[aps,prl,floats,graphicx,amssymb,epsf,twocolumn]{revtex}
\begin{document}
\draft \wideabs{
\title{
Peierls transition as a spatially inhomogeneous gap suppression}
\author{V.~Ya.~Pokrovskii, A.V.~Golovnya, and S.V.~Zaitsev-Zotov}
\address{Institute of Radioengineering and Electronics, Russian Academy of
Sciences, 103907 Moscow, Russia}
\date{\today}
\maketitle
\begin{abstract}
We propose a model of the Peierls transition (PT)
taking into account amplitude fluctuations
of the charge-density waves 
and spontaneous 
thermally activated suppression of the Peierls gap, akin to 
the phase slip process. 
The activation results in the exponential growth of the normal
phase with increasing temperature. 
The model fairly describes the behavior of resistance,
thermal expansion, Young modulus and specific heat 
both below and above the PT temperature $T_P$. 
The PT appears to
have a unique nature: it does not comprise $T_P$ as 
a parameter, and at the same time it has features of the
1st order transition. The possible basis for the model is
activation of
non-interacting amplitude solitons perturbing
large volumes around them.

\end{abstract}
}

Description of the Peierls transition (PT) in quasi 1-dimensional 
(1D) conductors still remains a controversial problem. The widely used
mean-field (MF) approach works poorly, firstly because of strong
1D fluctuations. {\it E.g.}, it predicts the PT
temperature $T_P$ much above the observed value. 
Say, for the typical compound TaS$_3$, the energy gap 
$2\Delta$ is  1600~K, and the MF value of $T_P$ should be
$2\Delta/3.5=460$~K, while the experimental value is 220~K.
The large
fluctuations reveal themselves well above $T_P$: the value of the pseudogap
is close to the low-temperature value  \cite{ItNa,gorshunov},
the threshold non-linear conduction \cite{MbIRC} indicates 
the charge-density wave (CDW) state within the fluctuating volumes.
At present only qualitative attempts to explain
these experimental facts are undertaken. 
Though certain success is achieved in fitting the behavior of different
values near $T_P$ \cite{Bribzor,BrillPRL,MozSyM}, the relations 
involved are semi-empirical, and their physical sense is not quite clear.
Another treatment of the PT is given by the generalized Ehrenfest relation
\cite{Ehr} 
between the specific heat, expansivity, Young modulus 
anomalies and stress-induced shift of the transition temperature. 
This relation works nice for some materials, {\it e.g.}
for K$_{0.3}$MoO$_3$ -
the blue bronze \cite{Bribzor},
but fails for others, like for TaS$_3$ \cite{BrillEPJB}.

The general approach to the CDWs is to consider them  
as a spatially homogeneous
state up to $T_P$. However, a recent theoretical study
has demonstrated that
thermal fluctuations of the CDW
stress may be very large at $T\approx T_P$, so that the r.m.s. shift of 
the chemical potential level from the middle-gap position
appears
comparable with $\Delta$ \cite{Art}, and one can expect temporal local
suppression of the gap.
So, it is natural to suppose 
that the
Peierls state in the vicinity of $T_P$ should be considered as a mixture 
of the Peierls phase (the CDW) and the state with suppressed gap.
Studies of noise have revealed spontaneous phase-slippage (PS)
process \cite{EPL,fluctran} in the vicinity of $T_P$, which also implies local 
temporal suppression of the Peierl gap.  
The PS is fairly described as a thermally activated process 
\cite{gillPS,borzzn86,rounding}, so it would be intriguing to
extend the aproach for the description of the PT.

In the present Letter
we propose a model in which the fluctuations of $\Delta$ are
phenomenologically 
introduced as thermally
activated local gap suppression (LGS).
With increasing $T$ this process gives rise to the activation 
growth of 
concentration of normal phase as $\exp(-W/T)$, $W \gg \Delta$.
The approach can be extrapolated above
the transition temperature. 
PT appears to be smeared out, but at the same time 
has features of the 1st order transition. The model appears to describe
fairly the behavior of various parameters near $T_P$. 
The plausible basics for the model is the excitation of amplitude solitons
perturbing considerable volumes around them. 

To introduce the fluctuations, we shall use the following relation 
giving the frequency of the LGS acts  
per unit volume (see also \cite{rounding}):

\begin{equation}\label{f}
f=f_a \exp(-W/T),
\end{equation}
where $f_a$ is an attempt frequency.
Here the essential point 
is that each LGS act 
results in
a temporal nucleation of the normal--state volume $v_0$ 
having certain lifetime
$\tau$. 
Then the fraction of the normal-state volume due to 
the spontaneous LGS process is 

\begin{equation}\label{normalv}
v=v_0\tau f.
\end{equation}
This fraction grows exponentially with increasing $T$. At high enough 
temperature $v$ becomes of the order of 1, so we should take into account
the shrinking of the Peierls--state volume. 
So, instead of (\ref{normalv}) we obtain:
$v=v_0\tau f (1-v)$,
or

\begin{equation}\label{normalvfin}
v=\frac{v_0\tau f}{1+v_0\tau f}.
\end{equation}
This relation, the principal one for our model, gives growth of $v$ 
from 0 to 1 with 
increasing $T$. The 
growth has the form of a step centered at
$T=W/\ln(v_0\tau f_a)$, at which $v=1/2$.
The step is
smeared out by $\approx T_P^2/W$ 
(or $1/W$ in the $1/T$-scale).   Extrapolation of the LGS
description above $T_P$ gives a way to treat the entire PT.
Within this approach the transition consists in 
the LGS-induced destruction of the CDW state. 

Let us compare the resulting relation~(\ref{normalvfin}) with the experiment. 
We shall refer mostly to TaS$_3$,--
a typical CDW compound, which is among the widely studied 
quasi 1D conductors \cite{mobzor}. 
The large ratio $2\Delta/T_P$ for TaS$_3$, 
as well as highly anisotropic structure
(the anisotropy of conductivity 
is about 100 at room temperature)
argue for the strong fluctuations of the CDW order parameter.

\begin{figure}[t]
\hskip -0.2cm
\includegraphics[scale=0.5]{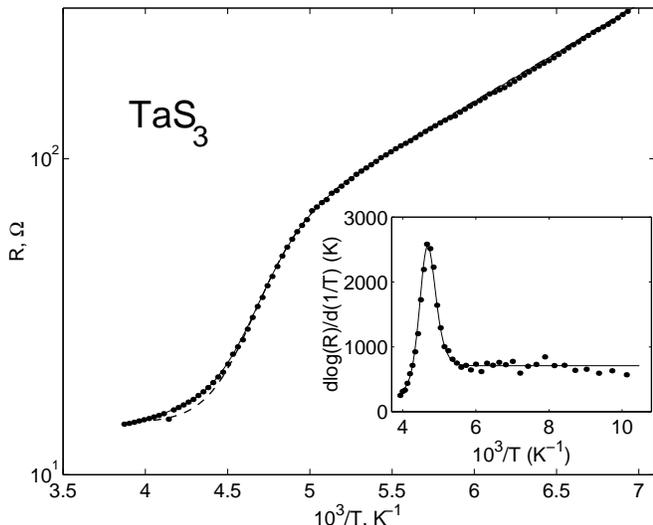}
\caption{Fits of a typical $R(T)$ curve for TaS$_3$ (points) with 
Eq.~(\ref{R}); for the solid line   
$W=6600$~K, $v_0\tau f_a=3 \cdot 10^{13}$, $\Delta=710$~K. The broken line
shows a similar fit with $B=0$. The {\it inset} shows 
the corresponding logarithmic derivatives.}
\label{FIG1}
\end{figure}

We are beginning with the temperature dependence of resistance, the most
common curve characterizing the PT. Within the
present model we should calculate the resistance of a mixture of two
phases with different resistivities, $\rho_c$ and $\rho_n$. We shall consider
the Peierls-state resistivity $\rho_c \propto \exp(\Delta/T)$, and the normal state 
resistivity $\rho_n=
A+BT$, where $A$ and $B$ are constants. 
To calculate the resulting resistivity one should consider
a complex electric connection of the domains of each phase. 
For simplicity, we take the contribution
of each phase to the resistivity to be just proportional to the volume 
fraction of the phase, as if for connection in series \cite{parallel}:

\begin{equation}\label{R}
\rho= v\rho_n+(1-v)\rho_c.
\end{equation}
Note, that the fluctations are known to contribute to the conductivity
of the Peierls phase due to thermal depinning of the CDW 
\cite{rounding,Gilround,thornround}. 
According to the model \cite{rounding}
this contribution is governed by the LGS as well and grows as $\exp(-W/T)$.
Here we shall not distinguish it from that of the normal phase. 

Fig.~1 presents an example of a fit of the 
$R(T)$ curve for TaS$_3$
with Eq.~(\ref{R}).
The fitting 
is splendid, but one should 
note that we have taken $B<0$, 
which is unreasonable for a metal. 
Even if we take $B=0$, {\it i.e.} 
$\rho_n =$~const, the fitting above $T_P$ becomes considerably worse (see
the broken line).
Below we shall introduce a modification of the model above $T_P$
explaining the slower drop of the CDW fluctuations.

Let us now probe the model for
the values which commonly characterize thermodynamic transitions, 
such as thermal 
expansion (TE), Young modulus ($Y$) and 
specific heat $c_p$ (see {\it e.g.} \cite{BrillPRL}).

\begin{figure}[t]
\includegraphics[scale=0.4]{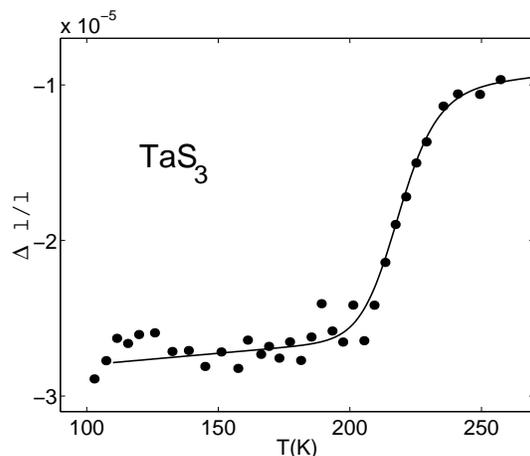}
\caption{$\delta l/l$ {\it vs.} $T$ for a TaS$_3$ sample. The background
approximated with a 2-nd order polynomial is subtracted. The solid line gives 
a fit with Eq.~(\ref{normalvfin}) with $W=6500$~K.}
\label{FIG2}
\end{figure}

To perform TE measurements, we have developed an interferometric 
technique for measurements of needle-like samples \cite{PRL,RSI}. 
Fig.~2  gives the temperature dependence of the relative length
change $\delta l/l$ for TaS$_3$ 
in the vicinity of the PT.
To exclude
the contribution of length hysteresis \cite{PRL}, we 
present the half-sum of the results obtained upon heating and cooling the
sample. A similar curve results
if we apply electric field exceeding
$E_T$ to remove metastability each time before measuring $l$. Evidently,
the curve presented 
is close to equilibrium. For all 
our measurements cooling below $T_P$ results in a {\it drop} of length by
about $10^{-5}$. This result is quantitatively similar to that
obtained for K$_{0.3}$MoO$_3$ in the in-chain direction \cite{mozurKTP}. 

TE at the PT has been discussed in \cite{mozurKTP}. {\it E.g.},
it could be
understood within the anharmonic model, taking into account the 
anharmonic effect of the lattice distortion associated with the CDW.
Without deepening into details, we just assume that the length increase
with heating
is proportional to the fraction of the normal phase. Thus, the
$l(T)$ step should be described by Eq.~(\ref{normalvfin}). The fit with 
$W=6500$~K 
is quite nice (Fig.~2, the solid line).

As another example we consider the Young modulus temperature 
dependence, $Y(T)$. 
One can expect a drop of $Y$ due to the same anharmonic effects.
So, with $T \to T_P$ from above one can expect
a decrease of $Y$ proportional to the 
fraction of the CDW volume. However, this is not the whole effect.

\begin{figure}[t]
\hskip -0.15cm
\includegraphics[scale=0.47]{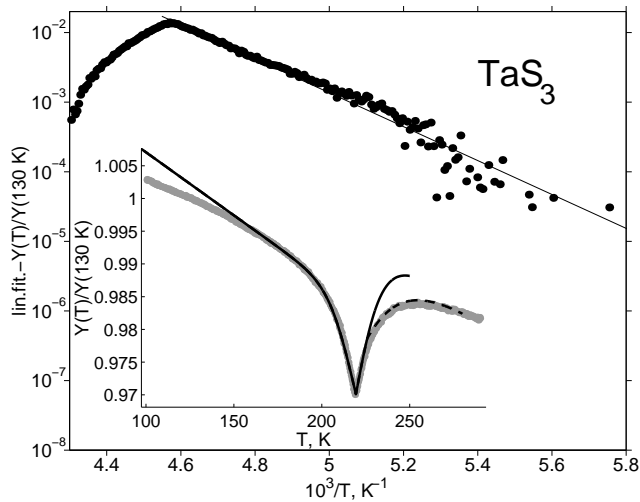}
\vskip 0.2 cm
\caption{Arrhenius plot of the deviation of $Y$ from a linear dependence 
$Y(T)$. The
slope of the solid line corresponds to $W=5600$~K. The {\it inset} shows the 
fit of the whole $Y(T)$ curve. The broken line takes into account that
for high $v$ ($T>T_P$) $1-v \propto \exp(-W/2T)$.
The data are taken from \protect\cite{MozSyM}.}
\label{FIG3}
\end{figure}

Let us recall that
depinning of the CDW below $T_P$ reduces the Young modulus 
\cite{Bribzor,Mozurkexp,Mozurkth}.
(This effect is associated with fast relaxation of
the CDW deformations which in the pinned state contribute to 
$Y$ \cite{Mozurkth}.)
As we mentioned above, with increasing $T$, the fluctuations result in
the spontaneous depinning of the CDW \cite{rounding,Gilround,thornround}, 
the concentration of the depinned state growing as $\exp(-W/T)$ \cite{rounding}, 
which results in the 
drop of $Y$ with approaching $T_P$ from below. Thus, a dip of $Y(T)$ is
expected at $T_P$ \cite{MakiYoung}. 
The value of the depinning drop of $Y$ depends
on the particular compound. Being anomalously strong for TaS$_3$ 
\cite{Mozurkexp}, it is not observed for 
the blue bronze, $\delta Y/Y<5\cdot10^{-5}$ \cite{27frombribzor,comment3D}. 
Thus, the dip in $Y(T)$ at $T \to T_P-0$ should be large for TaS$_3$,
and much weaker, if any, for the blue bronze.
This expectation agrees with the experiment: inset to Fig.~3 shows
$Y(T)$ for TaS$_3$ from \cite{MozSyM}. Large drop of $Y$ is seen
with $T$ approaching $T_P$ both from above and from below, whereas
only a small dip of $Y(T)$ at $T \to T_P$
from below is observed
for the blue bronze
in the in-chain direction \cite{Bribzor,BrillPRL}. 

Fig.~3 shows the dependence $\delta Y(T)$ below $T_P$ in the
Arrhenius axes (a linear dependence $Y(T)$ is subtracted). 
The slope of the solid line gives the activation
energy 5600~K, 
giving a good fit nearly up to $T_P$.
To fit the whole $Y(T)$ curve we present $Y$ as
$vY_n+v_pY_p+v_rY_r$, where the indices $n$, $p$ and $r$ refer to the
normal, pinned and relaxed states respectively 
($v+v_p+v_r=1$).
The drop of $Y$ in comparison with the normal state can be presented as
$\delta Y =(Y_r-Y_n)v_r+(Y_p-Y_n)v_p$, 
where $v_r=(1-v)\min(f_r \exp(-W/T),1)$, and $v_p=(1-v-v_r)$.
The inset to Fig.~3 shows the fit with $W=6000$~K, 
$f_r=7.8\cdot 10^{11}$, 
$v_0\tau f_a= 6 \cdot 10^{11}$, $(Y_r-Y_n)=0.0443$, $(Y_r-Y_p)=0.0111$ 
in the normalized units.
The fit is quite nice, but above $T_P$
the fluctuations fall down slower than the fit gives, as it could be 
expected (recall also Fig.~1). 

\begin{figure}[t]
\hskip 0.4cm
\includegraphics[scale=0.3]{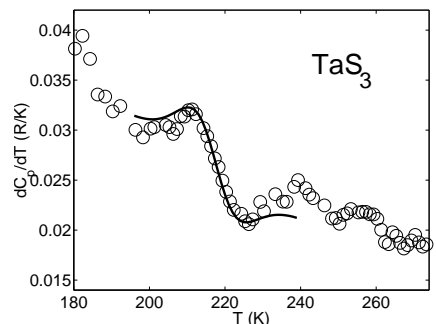}
\caption{The temperature derivative of the specific heat of TaS$_3$.
The fit is given by d$^2v/$d$T^2$ (Eq.~(\ref{normalvfin})) with $W=10000$~K.
The data are taken from \protect\cite{BrillEPJB}.}
\label{FIG4}
\end{figure}

It is clear from the examples above that the PT consists
in a gradual switching to the state as if having higher free
energy, and thus looks as a smeared--out 1st order transition.
The common check for the 1st order transition is the latent heat.
Because of the smearing out one can expect a maximum of specific heat $c_p$.
A cusp-like feature is clearly seen on the $c_p(T)$ curve for the
blue bronze \cite{BrillPRL}. Recently a similar feature has been observed
also for TaS$_3$ \cite{BrillEPJB}. Being very faint, it was 
detected as a zigzag pattern on the derivative d$c_p/$d$T$. Fig.~4
presents the data from \cite{BrillEPJB} together with 
d$^2v/$d$T^2$, Eq.~(\ref{normalvfin}) (the
background change of $c_p$ is approximated with a straight
line). It is clear that the form of the feature is at least 
approximately described by our model. Other words, the CDW
formation is accompanied by a smeared out step of latent heat
whose width is of the order of $T_P^2/W$. 
Note that the values of the 
latent heat 
$Q \approx 0.25~R \cdot $~K $ = 5 \cdot 10^4$~J/m$^3$ \cite{BrillEPJB}
($R$ is the universal gas constant) and the length change 
$\delta l/l =10^{-5}$  appear 
to be consistant
with the Clausius-Clapeyron equation 
${\rm d}T_p/{\rm d}\sigma = -T_p \frac{\delta l}{l} /Q$ if one takes 
${\rm d}Tp/{\rm d}\sigma \sim 1$~K/kbar \cite{mobzor,MozSyM} ($\sigma$ 
is the stress along the chains) \cite{Brillpriv}.

Above $T_P$ one should bear in mind the
small sizes of the remnant CDW volumes $v_c$. 
As soon as they shrink down 
below $v_0$, Eq.~(\ref{normalv}) is no longer 
valid, because the new normal volume 
due to the LGS cannot exceed $v_c$. 
With simple assumptions at high enough $T$ one can obtain 
$(1-v) \propto \exp(-W/2T)$; the appropriate 
fit for $Y(T)$ is given with a broken line in the inset to Fig.~3.
This consideration also explains the behavior of
the $R(T)$ curve above $T_P$ (Fig.~1). 

Thus, the LGS model fairly describes the temperature evolution 
of the principle
parameters in the vicinity of the PT. All the fits proposed
have transparent physical sense.
The main parameter
of the model, - the energy $W$, is close to the values for the
barrier characterizing thermally initiated 
PS \cite{gillPS,borzzn86,rounding}. Evidently, LGS is
governed by the same process as PS.
(Some indications  of the connection between PS and the 
PT have been given in the early works 
\cite{EPL,fluctran,rounding,NT}.) 
So, it would be natural to consider excitation 
of dislocation loops \cite{MakiPS} 
as a precursor effect below $T_P$.
Excitation
of the dislocation loops is being considered as a possible origin of
softening of solids \cite{loops} or similar transitions in liquid helium and
HTSCs \cite{vortices}. This approach gives critical expansion and 
proliferation of the loops due to their
mutual screening.
The apparent absence of the critical behavior in our case could mean that
the CDW excitations practically do not interact up to $T_P$. Note that
while for a conventional crystal
the smallest possible radius of a dislocation loop is 
of the order of the lattice constant, 
for the CDW it is of the order of $\xi_\bot$. 
Such an object ({\it i.e.} an
amplitude soliton) covers a volume 
$\sim \xi^3 \equiv \xi_\parallel \xi_\bot^2$, where $\xi_\parallel$  and 
$\xi_\bot$ are the in-chain and the transverse
amplitude correlation lengths respectively; the soliton
can perturb
a still higher volume, where, say, the conductivity is
increased.
Thus, the condition $v=1/2$ can be achieved 
when the 
concentration of the excitations is still $\ll 1/\lambda s$, where
$\lambda$ is the CDW wavelength, and $s$ is the area per chain.

The concentration of the solitons could be estimated as
$\frac{1}{\lambda s} \exp (-W/T)$, 
where $W$ is the energy of such an excitation.
Then, $v \sim \frac{\xi^3}{\lambda s} \exp (-W/T)$. 
At $T=T_P$ we have 
$\frac{\xi^3}{\lambda^3} \exp (-W/T) \approx 1$, and
come to the estimate:

\begin{equation}\label{TPENP}
T_P \simeq \frac{W}{\ln(\xi^3/\lambda s)}
\end{equation}
With $\xi^3/\lambda s=10^3$ we obtain $T_P=W/7$, which
can give an idea of the low value of $T_P$ in comparison with $W$. 
A higher ratio $W/T_P$ might be obtained if we take into consideration 
the large wavelengths of the fluctuations of the CDW stress along the
chains. According to
\cite{Art} they can considerably exceed $\xi_\parallel$ and, consequently, the
LGS volumes could appear much larger than $\xi^3$.


According to our estimates, 
the excitations
would begin to turn into dislocation loops, only
above $T_P$. 
So, within the model, the metallic state develops at lower
temperature than the critical behavior is expected to begin. 
As
far as we comprehend, the approach proposes a new type of phase
transition, which does not comprise $T_P$ as 
a parameter. 
At the same time, the PT in a sense resembles a
1st order transition.
The model successfully works both below and above $T_P$, though further
extrapolation of the approach to higher temperatures requires 
further development of the model. The underlying microscopic
mechanisms of the LGS also needs deeper understanding. Though
the model requires further grounds 
it gives a limpid
insight into the processes inside the CDW near $T_P$.

We are thankful to J.W.~Brill, G.~Mozurkewich  and
D.~Stare\v{s}ini\'c for granting the experimental results at our disposal,
to V.V.~Frolov and R.E.~Thorne for furnishing the samples,
to V.I.~Anisimkin, L. Burakovsky and  S.N.~Artemenko
for helpful discussions. This work has been supported by the RFBR
(grants No 01-02-17771, 02-02-17301), 
Jumelage (CNRS and RFBR), INTAS (grant No 01-0474)
and by the State programme ``Physics
of Solid-State nanostructures''.


\begin{thebibliography}{8}
\bibitem{parallel}If instead of resistivities we sum up 
conductivities (connection in parallel), the result is similar. 
Actually, one can expect a weak criticality associated with
the current percolation through the normal-phase 
cluster, but this concerns only
the transport properties.
\bibitem{comment3D}To our
understanding \protect\cite{PRL}, the effect is due to 3D features of
the electronic structure, which are individual for each compound.
\end{thebibliography}
\end{document}